# Scope of Magnetic Tunnel Junction Based Molecular Electronics and Spintronics Devices

**Pawan Tyagi [1,*], Edwards Friebe[1], and Collin Baker [1]**

[1] Mechanical Engineering Department, University of the District of Columbia 4200 Connecticut Av. NW, Washington DC 20008; E-Mail: ptyagi@udc.edu ;

**Abstract:** Dream of developing molecule-based logic and memory device is >70-year-old. Presently, molecule-based devices are also considered for quantum computation hardware; recent study have shown the molecule connected to metal leads can perform qubit-based logic operation. This is an interesting question why experimental progress is still very slow even when scope of molecule-based devices may govern the advancement of the next generation's logic and memory devices for the highest possible computer technologies. Molecules have the potential to be unmatched device elements as chemists can mass produce an endless variety of molecules with novel optical, magnetic, and charge transport characteristics. However, the biggest challenge is to connect two metal leads to a target molecule(s) and develop a robust and versatile device fabrication technology that can be adopted for commercial scale mass production. This paper discusses distinct advantages of utilizing commercially successful magnetic tunnel junctions as a vehicle for developing molecular electronics and molecular spintronics devices. We describe the use of a tunnel junction with the exposed sides as a testbed for molecular devices. On the exposed sides of a tunnel junction molecules are bridged across an insulator by chemically bonding with the two metal electrodes; sequential growth of metal-insulator-metal layers ensures that separation between two metal electrodes is controlled by the insulator thickness to the molecular device length scale. This paper critically analyzes and discusses various attributes of tunnel junction based molecular devices with ferromagnetic electrodes for making molecular spintronics devices. Here we also strongly emphasize a need for close collaboration between chemists and magnetic tunnel junction researchers. Such partnerships will have a strong potential to develop tunnel junction based molecular devices for the futuristic areas such as memory devices, magnetic metamaterials, and high sensitivity multi-chemical biosensors etc.

---

## 1. Introduction

Molecule based devices may lead to the next generation of logic and memory devices [1-4]. Molecular devices can utilize charge and spin properties of electrons and hence can be considered as the most promising candidates for replacing old technologies and producing novel forms of devices [4]. A molecular device using single or few molecules have potential for creating a high-density device architecture with low power consumption and high speed [5]. Molecular electronics and spintronics can be considered as a potential route to cross the fast approaching miniaturization limit for the conventional semiconductor devices. Molecular devices, depending on their design, can use a small ensembles or even individual molecules as functional building blocks in electronic and spintronics device circuitry [6]. Molecules can be unmatched device elements; chemists can mass produce an endless variety of molecules with exotic optical, magnetic, and charge transport characteristics [6, 7]. Due to configurable internal molecular structure, molecules may provide novel intrinsic functionality that is not found in the present day silicon electronics [8] and multilayer magnetoresistance devices [9]. With skills and resources for preparing molecular devices modern device engineers and researchers will not be limited to silicon or only a handful of material choices while solving challenging problems of finding advanced devices. In addition, molecule based devices have the potential to be more economical since a self-assembly process can be used to incorporate molecule into device form. However, developing a reproducible, commercially viable, long lasting, and economical molecular device fabrication technology continues to be a major impediment [4, 5].

Most of the molecular device fabrication approaches [5, 10, 11] attempted so far can be categorized under the following four groups [3, 4]. (i) Molecules sandwiched between a conducting film and scanning tunneling microscope(STM)/conducting probe atomic force microscope (CPAFM) tip (Figure 1a) [12], (ii) molecular monolayer sandwiched between two conducting electrodes (Figure 1b), [5, 13] (iii) molecule placed in a nanoscale gap of a metal break junction (Figure 1c) [14-16], (iv) molecules chemically bonded to the two metal films of a tunnel junction along the exposed edges[3, 17-20]; this approach utilizes a vertical edge of thin film [21, 22] (Figure 1d-e). The first three molecular device fabrication approaches have been widely experimented for nearly a decade, but, failed to get matured to the extent of becoming commercially viable. This paper aims to highlight the distinctive advantages of the fourth approach that utilizes tunnel junction based molecular devices (TJMDs) for developing commercially viable molecular devices (Figure 1 d-e)[3]. The following is a brief overview of different molecular device fabrication approaches and subsequently we highlight the attributes of TJMDs.

## 2. Overview of molecular device fabrication schemes

The first molecular device approach utilizes STM/CPAFM probes to establish electrical contacts with one or few molecules out of a self-assembled monolayer on a conducting substrate (Figure 1a) [23-25]. The STM/CPAFM probes can interact with a molecule via space or via chemical bonding. The STM/CPAFM has been widely utilized to study the conduction properties of a single molecule. STM/CPAFM has also been mainly useful in deciphering molecular spectra with regards to metallic Fermi level. Seo et al. [23] studied structural phase dependency of conductance across thiolate self-assembled monolayers (SAMs) in different molecular junctions. Voss et al. determined the band gap of $Mn_{12}$ single molecular magnet [26]. This approach is good to study science pertinent to molecular conduction [23, 24], but not for a large scale device production. This approach suffers from a number of limitations: (i) due to bulky and challenging hardware requirement this approach cannot be used for mass production; (ii) it is extremely challenging to maintain a reproducible molecular channel length during the charge conduction study [24], (iii) The science and art of establishing reproducible and enduring molecule-metal contact using a STM/CPAFM tip is not well developed [24]; (iv) a STM/CPAFM tip based molecular devices utilizes mechanically unsupported molecule(s) and, therefore are very sensitive to surrounding noise that can interfere with device functionality [12, 27]. (v) a STM/CPAFM tip based approach has negligible potential to include ferromagnetic films or multilayers into circuitry; this approach has negligible scope for developing molecular spintronics devices.

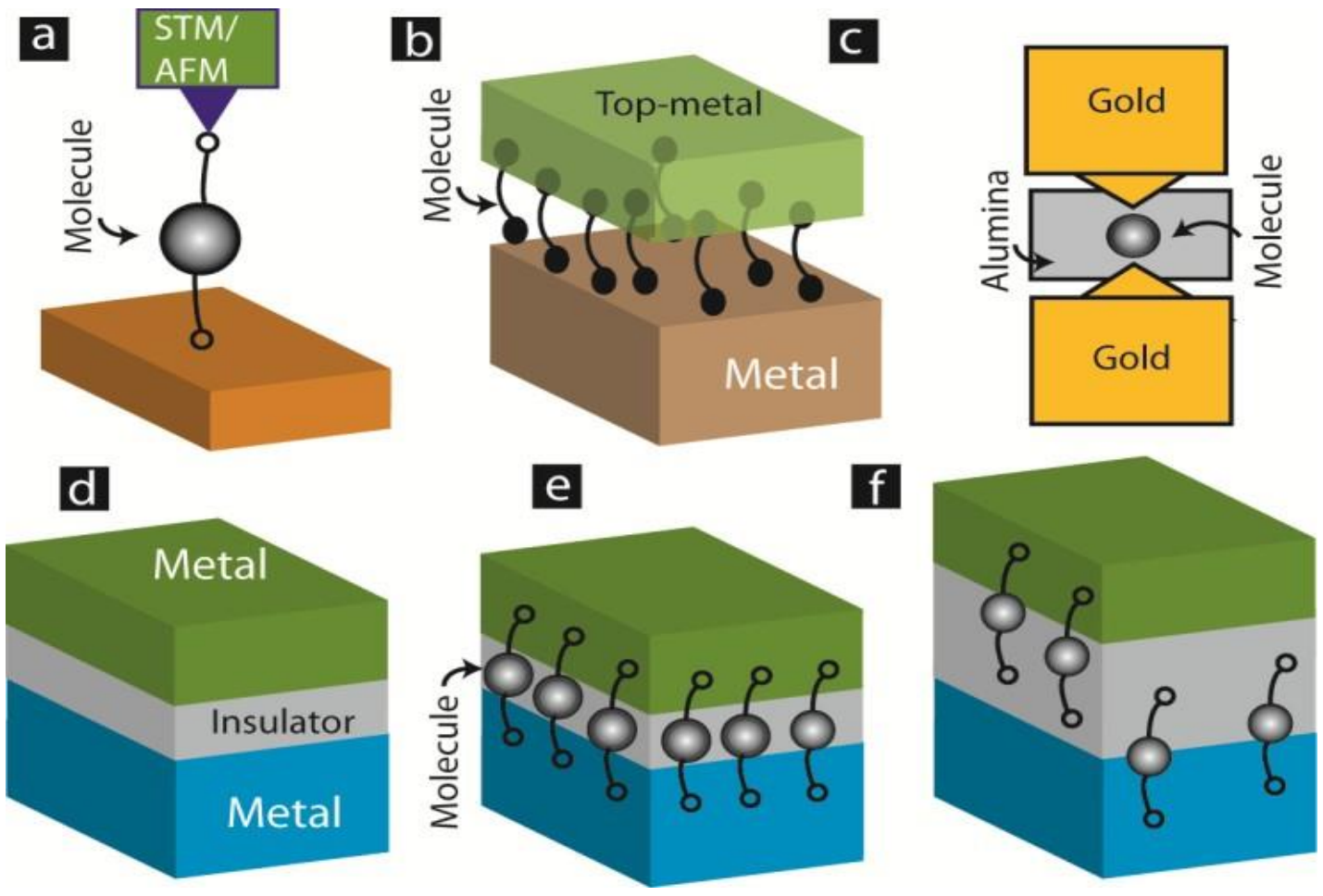

**Figure 1.** Molecular device designs: (a) molecule electrically connected to conducting substrate and to the tip of STM or CPAFM. (b) Molecules sandwiched between two metallic films. (c) Molecule in the gap of a metallic break junction. Tunnel junction (d) before and (e) after the molecule attachment on to the exposed edges. (f) High tunnel barrier thickness will disable molecules bridging.

The second molecular electronic device approach is based on sandwiching a molecular monolayer, comprising few thousands to millions of molecules between two metal films [5, 13]. The biggest highlight of this approach is the simplicity of the device concept (Figure 1B); however, in reality it is extremely challenging to develop robust and long-lasting devices with different metal electrodes [5, 28]. Recently, highly stable molecular devices were developed using conducting organics electrodes [28]; however, the scope of organic electrode based molecular devices is limited. The following unresolved demerits severely limit the scope of 2D molecular monolayer based device architecture: (i) a non-rigid molecular monolayer, whose primary purpose is to conduct charge or spin, is forced to serve as a mechanical barrier between two electrodes. The denseness of molecules in the molecular monolayer strongly depends on the quality of monolayer coverage on the bottom metal electrode [5, 11]. Generally gold (Au), which provides a reasonably good monolayer quality, is widely utilized as the bottom metal electrode [29]. However, other metal electrodes may lead to very different monolayer quality and may undergo oxidation. (ii) Time-dependent defects evolution in the molecular monolayer deteriorates the molecular junctions [4]. Metallic atoms from the electrodes can easily diffuse through the molecular monolayer to create a short circuit [30] . (iii) Only a limited variety of molecules can be used. Molecules, with a number of functional groups, side chains, and bulky centers like single molecular magnets [9, 26] will presumably form a less dense monolayer. This less packed layer may possess a lot of holes and penetrable space for the electrodes' atoms to diffuse towards each other and create short circuit; simple 12 carbon alkane molecules takes around ~2 $nm^2$/molecule in a monolayer [31, 32]. (iv) During the top metal electrode deposition the molecular monolayer may get damaged or distorted due to the interaction with energetic metal atoms [4]. (v) In this approach a large ensemble of molecules simultaneously contributes in the conduction process. This large ensemble of molecules may have a variety of molecular conformation and effective lengths [23, 27]; there is no reliable method to understand what the true representative molecular configuration and length in a completed device is. (vi) An arbitrary selection of metal electrode, such as ferromagnetic electrode [33], is also quite tedious with this approach [13]. Deposition of ferromagnetic electrodes on the molecular monolayer or depositing a high quality monolayer on ferromagnetic electrodes is quite challenging [13]. (vii) It is impossible to reverse the effect of molecules. Once a device is made one has to deal with molecular monolayer and unknown amount of defects within. It is extremely challenging to differentiate the role of defects in overall conductance.

The third molecular electronic device approach utilizes a planar metal-break junction with ~1-2 nm separation between the two metal electrodes [34]. The nanogap in a metal-break junction is bridged by one or few molecular channels (Figure 1f) [16]. This approach has been highly insightful in revealing a number of exotic phenomenon associated with the molecular conduction processes [14, 16, 35]. For instance, phenomenon like coulomb blockade, Kondo effect [36], transition between conduction modes [36], and the current suppression [37] are to name a few [3, 4]. A new class of molecules with tunable spin state, such as single molecular magnets, have been studied with these metal break-junctions [37]. However, break-junction based molecular devices suffer with numerous demerits making this approach narrow in scope. (i) Nonetheless, the fabrication of metal break-junctions necessitates the utilization of sophisticated instrumentation and generally exhibits a low yield [14]. (ii) Atomic level defects produced during the creation of a break-junction can mimic the molecular characteristics. For instance, in one study metal break-junctions exhibited Kondo effect even without inserting any molecule within the gap [38]. This observation was believed to arise from atomic artifacts. (iv) In most of the cases metal break-junctions were not characterized without molecular channels; due to this reason a contribution or influence of atomic defect(s) in the molecular conduction process is difficult to identify [14, 38]. (iv) This approach typically thrives on the Au, which easily produces a nanogap after an electromigration step [14]; rarely other metals have been used for producing break-junctions [14]. (vi) Break junction based devices have not demonstrated the ability to undo the effect of molecular conduction channels and retrieve the properties of the break junction testbed. This experiment focusing on reversing molecules effect can be used as a litmus test to assure that the molecular device channel is the only viable conduction bridge.

The three widely attempted molecular device fabrication approaches (Figure 1a-c) have limited scope. These approaches need to overcome a large number of technological hurdles to be a viable option for the commercial production of novel molecular devices. In addition, the above mentioned three approaches may require the development of a new commercial scale device fabrication protocol, instrumentations, and laboratory etc. Recently, several research groups independently attempted to develop tunnel junction based molecular device (TJMD) [3]. This approach has strong potential to overcome many limitations discussed in the context of other molecular device architectures. The following sections elaborate the philosophy and significant attributes of TJMDs.

## 3. Tunnel junction based molecular devices

A TJMD focuses on utilizing a molecular device element only as a medium of charge and spin transport between the two metal electrodes [4, 33, 39]. A TJMD fabrication protocol involve two key steps: (i) fabrication of a tunnel junction with exposed side edges where the maximum distance between two metal electrodes is equal to the thickness of tunnel barrier (Figure 1d) [3], and (ii) chemically bonding the molecule(s) of interests across the insulator gap (Figure 1e) [18, 19]. The following are the distinct advantages and highlights of this approach.

### *3.1 Molecules serve as a device element not as a physical barrier*

The main attribute of a TJMD is that it does not use molecule(s) as a mechanical spacer between the two electrodes [19]. An ultrathin insulating spacer such as alumina (AlOx) [18], silicon dioxide ($SiO_2$) [19], etc. serves as a robust physical spacer keeping metal electrodes at the distance that is slightly smaller than the molecule length (Figure 1e and Figure 2a). This attribute addresses the three key issues observed with molecular device scheme based on sandwiching a molecular self-assembly between the two metal electrodes (Figure 1b). TJMD's insulating spacer prevents the propagation of the top metal electrode's atom reaching to the bottom electrode during deposition process, which is quite likely with the molecular device architecture utilizing molecules as a mechanical spacer (Figure 1b) [40]. The TJMD's insulating spacer also serves as a robust diffusion barrier and hence minimizes the failure due to the diffusion of metal atoms via spacer [20]. A TJMD is also quite forgiving towards the quality of molecular self-assembly between the two metal electrodes [40]. Since molecules of interest always stay on edges (Figure 1e) hence, an imperfect molecular self-assembly [41] will still work fine and will provide a functional TJMD.

In a TJMD, molecular channels are chemically anchored between the two metal electrodes and hence presumably attain a stable effective length (Figure 1e). Molecular channels are also mechanically supported by the insulating spacer and metal leads. Due to this attribute, a molecular conduction channel is expected to exhibit rather consistent charge and spin transport characteristics; molecules' transport characteristics are very much dependent on their interaction with surroundings and configuration [42]. It must be noted that the STM/CPAFM tip based devices exhibited varied moleculartransport characteristics [43]. It is extremely challenging to force a mechanically unsupported molecule with a variable effective length to demonstrate a consistent charge and spin transport characteristics [43, 44]. It is also noteworthy that break-junction based molecular devices also have similar advantages as compared to the TJMDC approach. A break junction also mechanically supports molecular conduction channel(s). However, the biggest difference is in the method of creating an insulating spacer between the two electrodes. The TJMD's insulating spacer is created by highly reproducible and high yield thin film deposition process route that is accomplished with economical microfabrication instruments [3]. In break-junctions an insulating spacer is produced by a poorly reproducible and low yield elctromigration step which is accomplished with expensive nanofabrication tools [16, 34].

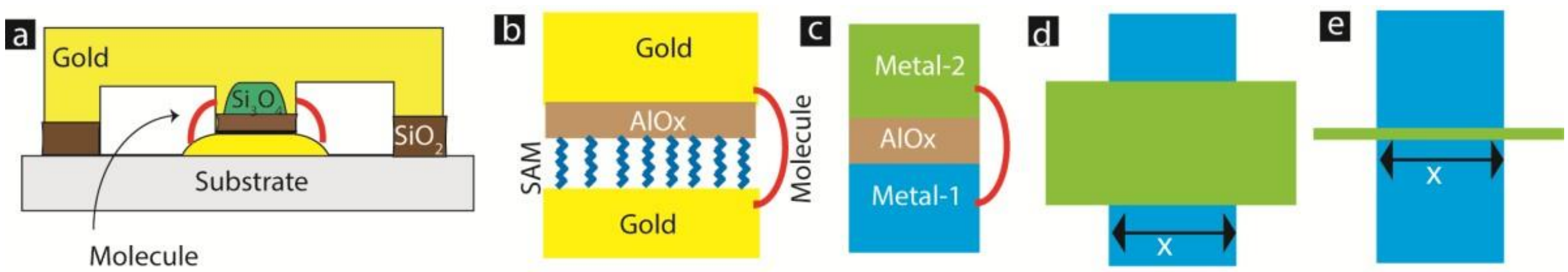

**Figure 2.** Strategies for reducing background conduction in TJMDs: (a) mushroom shaped double insulator spacer containing flat $SiO_2$ and $Si_3N_4$ between two gold electrodes (adopted from [ 1 9 ] ) . (b) Double insulator spacer containing self-assembled monolayer (SAM) and AlOx used between the two Au electrodes (adopted from [45]). (c) Single layer ~2 nm AlOx optimized according to TJMD electrode [46]. Reducing leakage current by reducing the junction area from (d) large cross section to (e) small cross section by keeping the exposed edge length the same.

Utilizing a separate and an adjustable insulator as a spacer between the two metal electrodes also enables a unique and very useful control experiment under the TJMD approach. One can set the physical thickness of the insulator spacer to be smaller than (Figure 2a) or more than (Figure 2c) the effective length of molecular channels to demonstrate that only intended molecular bridges play the role in enhancing conduction [18] or affecting the overall device behavior [33].

*3.2 Ability to tailor molecular current with respect to leakage current*

Observing the effect of molecular conduction channels in the TJMD approach depends on the leakage current through the tunnel barrier. The utilization of a bilayer insulator can reduce the leakage current. Ashwell et al. [19, 47] incorporated a mushroom shape bilayer tunnel barrier (Figure 2a), comprising of silicon nitride ($Si_3N_4$) and silicon dioxide ($SiO_2$) insulators, to reduce the leakage current via the tunnel barrier. The mushroom shaped insulator provided a relatively small gap between the metal electrodes along the edges; however, towards the center the effective tunnel barrier thickness was significantly higher as compared to the tunnel barrier thickness close the edges [47]. Hu et al. [45] used a combination of self-assembled monolayer (SAM) of organic molecules and alumina (AlOx) insulator to reduce the leakage current. Tyagi et al. [18] used a single layer of AlOx insulator; however, significant efforts were made to optimize the quality of ultrathin AlOx for the electrode materials of interest [46]. Dedicated studies revealed that the leakage current through ~2 nm AlOx was strongly dependent on the bottom metal electrode [48]. For instance, the AlOx deposition recipe optimized for the NiFe electrode showed a lesser leakage current and higher break down voltage on Ta bottom electrode but, a higher leakage current and lower break down voltage on copper(Cu) bottom electrode [48].

We recommend reducing the leakage current by reducing the width of the top electrode to reduce the cross sectional area of a tunnel junction test bed (Figure 2d-e). One can keep the bottom electrode width quite wide to still provide plenty of sites for the molecular bridges along the edges. However, this scheme may only work with the liftoff based TJMD approach [39]. Tyagi et al. [18, 39] produced TJMDs with a typical tunnel junction of ~25 μm² area; overall ~10 μm length was available for posting molecular bridges along the tunnel junction edges [18, 33]. Reducing width of the top metal electrode to 1 μm or less will reduce the tunnel junction leakage current by a factor of five, making the molecule effect more prominent. Reduction in tunnel junction area will also prolong the TJMD's stability and lifetime. Chen et al. [17] have already demonstrated the use of e-beam lithography for producing a tunnel junction area less than 1 μm² [49]. However, one needs to be careful about the optimization of tunnel junction quality.

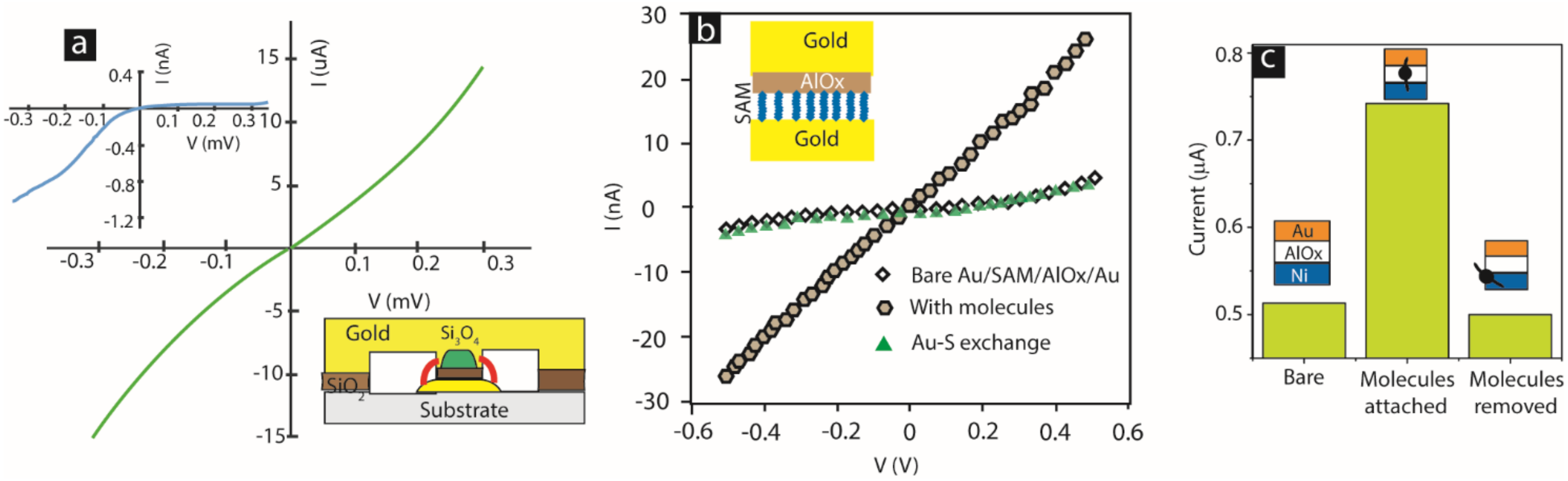

**Figure 3.** (a) Double insulator based TJMD showed reversing of current after breaking molecular conduction channel in acidified acetone (adopted from ref. [19] ), inset graph show current-voltage response from a bare tunnel junction. (b) SAM and AlOx insulator based tunnel junction (adopted from ref. [45]) and (c) Ni-AlOx-Au tunnel junctions showed current reversal due to Au-S exchange (adopted from ref. [18]).

### *3.3 Ability to perform reversible experiment*

All the molecular device testbeds are likely to possess artifacts and defects during fabrication or with time [3, 11]. The TJMD approach is also likely to encounter defects that are present within the tunnel barrier at the time of fabrication, or emerges with time due to stresses and diffusion of electrode materials through the insulator's weak points [46]. It is crucial to check if a molecular device behavior is due to the molecules only, not due to the artifacts. To do so one can attempt to reverse the molecule effect by disconnecting molecular device channels from metal electrodes and regaining the original transport characteristics of a tunnel junction testbed [3]. Since in TJMDs molecules are present at the exposed edges it is straightforward to remove or break the molecular bridges by the suitable chemical reactions or oxygen plasma [33]. So far four different studies demonstrated the reversal of molecular channels effect on a tunnel junction testbed [18, 19, 33, 45]. Ashwell et al. [19] produced bilayer insulator based TJMDs with Au electrodes by establishing molecular channels by modular chemistry between the two metal electrodes (Figure 3a). After the creation of complete molecular bridges a bare tunnel junction showed a clear enhancement in the current (Figure 2a and the inset graph). However, submersing a complete TJMD into the acidified acetone solution made molecule induced current enhancement disappear (Figure 3a); acidified acetone solution broke the Au-thiolate bonds [19]. Another study successfully demonstrated the reversal of molecule effect on the charge transport characteristics of Au/SAM/AlOx/Au configuration based TJMD [45]. Exchange reaction broke the Au-sulfur(S) linkage with molecular conduction channel (Figure 3b). Tyagi et al. [18] demonstrated the reversing of molecule effect on nickel(Ni)/AlOx/Au tunnel junction test bed using the mass action reaction to selectively break the Au-thiol linkage of the molecular channels (Figure 3c). After the breaking of Au-molecule bonding the Ni-AlOx-Au tunnel junction testbed current level became very close to the bare tunnel junction current. This study showed that Au-thiol bond was relatively weak with respect to Ni-thiol bond [18].

A molecule effect reversal could also produce original magnetic behavior of bare magnetic tunnel junction. An ensemble of several thousands of Ta/Co/NiFe/AlOx/NiFe magnetic tunnel junctions [33] with organometallic molecular clusters (OMCs) [50] were treated with oxygen plasma. Oxygen plasma burnt several molecular bridges along the exposed edges. Ferromagnetic resonance studies performed on a bare Ta/Co/NiFe/AlOx/NiFe were similar to resonance studies observed after the oxygen plasma treatment [33]. Use of oxygen plasma was rather harsh and risky for the single TJMDs used for transport studies [18]. It is noteworthy that other than the TJMD approach no other molecular device fabrication approach was demonstrated to have this important capability.

### *3.4 Avoiding inclusion of impurities during molecular self-assembly*

Making electrical connections with a target molecule to yield electrode-molecule(s)-electrode configuration is practicably straightforward with TJMDs. The attachment of molecular device elements to the metal electrodes is performed on a prefabricated tunnel junction [3, 18, 33]. A molecule can be attached to metal electrodes simply by submersing a tunnel junction into a molecular solution for some duration [19, 51, 52] or using electrochemistry [18, 49] to enhance the molecular self-assembly. It is noteworthy that during the process of molecular self-assembly, only those molecules which successfully bridged between the two metal electrodes serve as the molecular device elements. Remaining molecules stay on the electrode surface; without interfering with the molecular conduction process [18]. It must be noted that the space between the two electrodes is protected by the insulating layer and virtually no external impurities or artifacts can intrude inside the spacer [3]. Molecular device elements in a TJMD form a single molecule thick and one molecule wide ribbon along the exposed edge. In this configuration intermolecular interaction is presumably much simpler and less pronounced as compared to the intermolecular interaction in a self-assembled monolayer; the intermolecular interaction can strongly influence overall molecular conduction [27].

The TJMD is also tolerant towards the quality of a molecular self-assembly along the edges; even though molecules are not closely packed along the edges of the tunnel junction testbeds a TJMD will function satisfactorily. Due to this attribute a TJMD can employ a wide variety of metals and semiconductors; use of these metals and semiconductor may not be possible for othe molecular device fabrication approaches as the may lead to a defective 2D molecular self-assembly. For instance, in the scheme where molecules are sandwiched between the two metallic electrodes a large number of molecular configurations and pinholes will take part in the molecular conduction process [40]; molecular self-assembly will have convoluted and straight molecules along with other defects arising from the metal electrodes to impact the molecular transport. In addition, this approach is highly sensitive towards the quality of molecular self-assembly [32]; the pinholes in the molecular layer will create a short-circuit between the electrodes and lead to device failure [53]. In the case of break junction based molecular device, molecules are placed between the electrodes during the creation of a nm scale gap by electromigration. Electromigration is an energy intensive process which may not only damage or modify the molecules of interest but also can create additional defects [54]. The break junction devices, without molecule, were shown to exhibit Kondo resonance due to the artifacts [38]; Kondo resonance was also shown by the break junction devices due to the molecules [14]. One can easily observe that other molecular electrode fabrication schemes as discussed in section 2 are likely to have a number of artifacts; their application in futuristic molecule-based electronic and spintronics devices will be challenging.

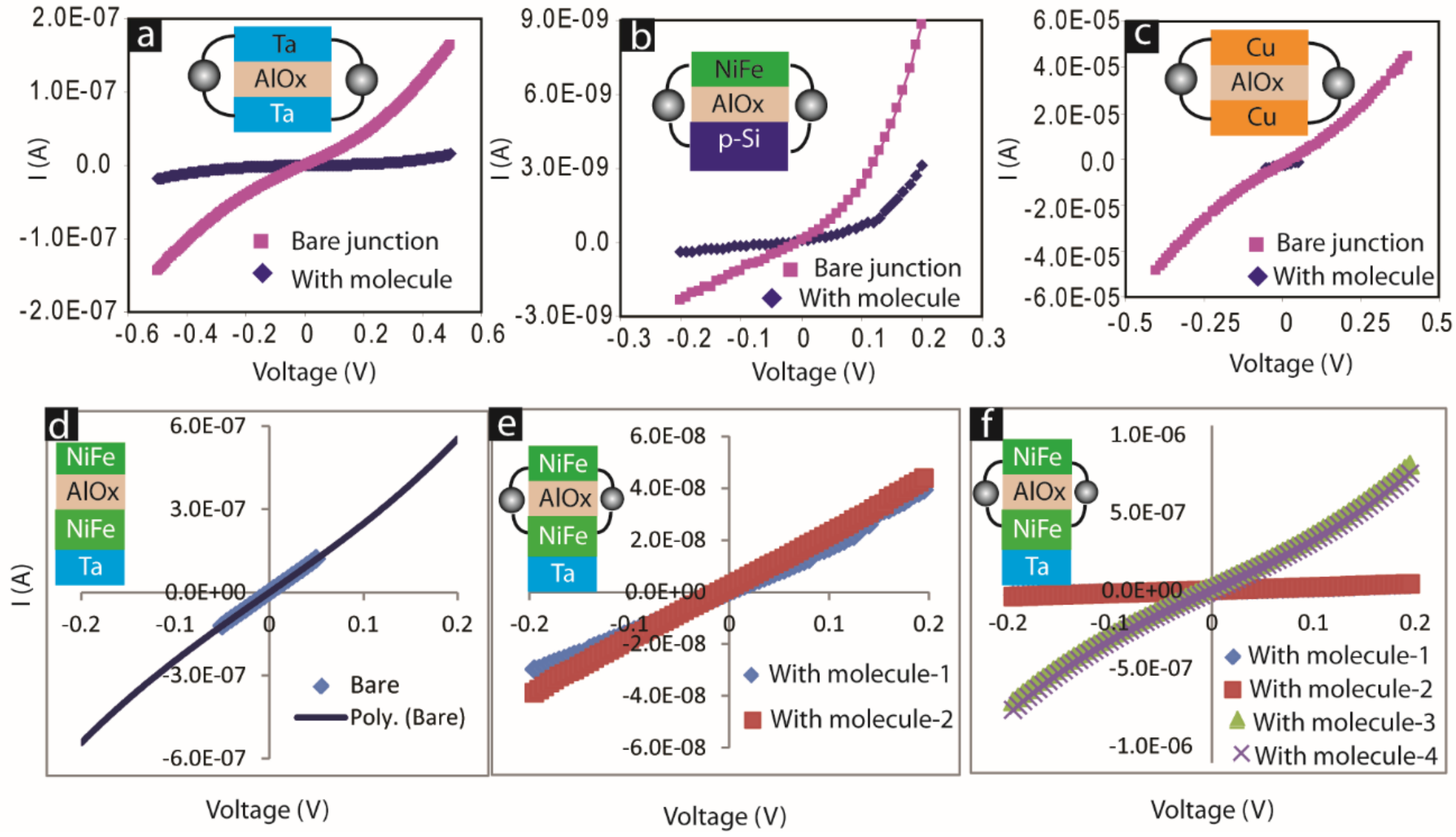

**Figure 4.** TJMD with (a) top and bottom Ta electrodes, (b) p-Si bottom and NiFe top, (c) top and bottom Cu electrodes. Note frequently, a bare junction current was recorded for the low voltage range to ensure long term device stability; low voltage range tunneling current can be extrapolated for high voltage range as shown. (d) TJMD with Ta/NiFe and NiFe ferromagnetic electrodes, black line is produced by fitting 3[rd] order polynomial on low voltage range data. (e) The appearance of temporary current reduction below leakage current level. (f) Repeating current-voltage measurement brought the TJMD current into higher current stage.

*3.5 Ability to arbitrarily choose metal electrodes*

A TJMD approach enables the utilization of different type of metallic electrodes, even in the same molecular device [18, 48, 55]. This capability makes a TJMD much superior as compared to other molecular device fabrication schemes [11]. The future of molecular spintronics devices may especially depend on how easily a molecular device approach can incorporate two magnetic electrodes of different magnetic properties. This difference in magnetic attributes is crucial to observe magneto resistance behavior [56] and intriguing phenomenon like Kondo level splitting [14]. One of the authors has contributed in the development of several TJMDs with different types of electrode materials (Figure 4a-d) [48, 55]. A Ta/AlOx/Ta tunnel junction produced clear increase in overall current with organometallic molecular clusters (OMCs) (Figure 4a) [18]. In-depth details of OMC synthesis and characterizations are published elsewhere [50, 57]. A tunnel junction testbed with p type silicon (p-Si) on the bottom side and NiFe on the top showed noticeable increase in the current after hosting OMCs channels between Si and NiFe along the edges(Figure 4b) [48]. Similarly, a copper (Cu)/AlOx/Cu tunnel junction showed an increase in current due to OMC channels (Figure 4c). Another device configuration, where Ni/AlOx/Au tunnel junction was used as a test bed [18], showed a clear increase in the current (Figure 3c) [18].

A number of TJMDs with two ferromagnetic electrodes showed intriguing phenomenon. A TJMD with Ta(2nm)/NiFe (10 nm) as the bottom electrode and NiFe (10 nm) as the top electrode (Figure 4d) showed a reduction in overall current below the leakage level of the tunnel barrier (Figure 4e). The fabrication protocol for this device is published elsewhere [18]. Both Ta and NiFe metals were determined to be highly chemical etch resistant when exposed to molecular solution [58]. In addition, NiFe was also resistant to low temperature oxidation [59]. This TJMD with two NiFe electrodes, abutted to a AlOx barrier, returned to a higher current state after the repetition of current-voltage studies (Figure 4f).

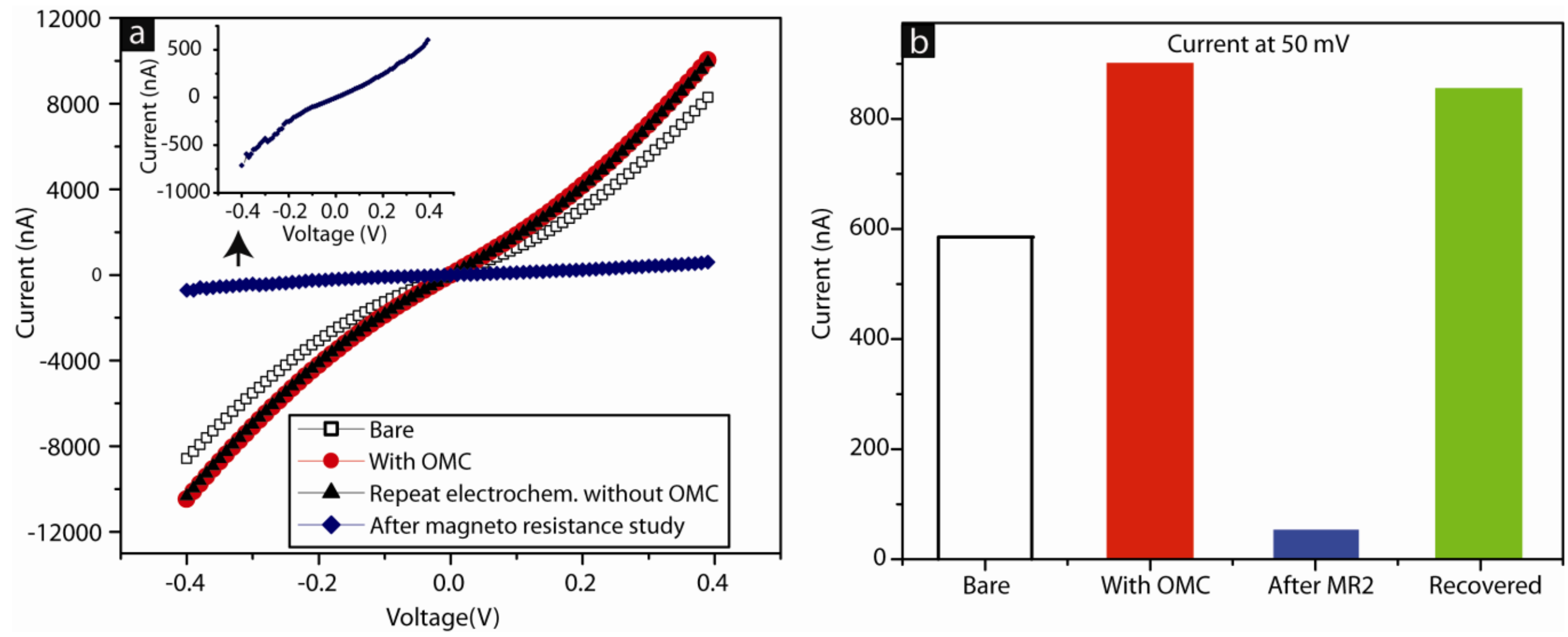

**Figure 5.** (a) The TJMD with Co/NiFe/AlOx/NiFe configuration showing current suppression during magneto resistance (MR) study; initially OMCs produced a stable current increase over the MTJ leakage level. (b) Net current at 50 mV in various stages of this TJMD shown in (a).

Interestingly, a TJMD with a Cobalt (Co)/NiFe bilayer bottom electrode and a NiFe top electrode exhibited significantly stable current suppression [60, 61] . According to magnetic studies the Co/NiFe and NiFe electrodes possessed different magnetic attributes [33]. For these studies OMCs [50, 57] served as molecular channels. OMCs produced a stable current increase when bridged along the edges of a (Co (5-7 nm)/NiFe (3-5 nm)/AlOx (2 nm)/NiFe (10 nm) tunnel junction (Figure 5a). Varying the Co thickness in 5-7 nm range and the NiFe film thickness in 3-5 nm range kept showing OMCs induced current suppression; overall ferromagnetic electrode thickness in the bottom electrode was maintained to be ~10 nm. To ensure that OMC channels, not the serendipitous defects arising from the OMC attachment protocol, produced this change a number of control experiment were performed. In one control experiment electrochemical protocol for molecule-metal bonding [18] was repeated with solvent only, without OMCs dissolved. This blank electrochemical step did not affect the magnitude of current achieved after the molecule attachment (Figure 5a).  After checking the stability of an OMC enhanced conduction the magneto resistance (MR) study was performed. The MR study measures TJMD current at fixed voltage with a variation of magnetic field. During the MR study this sample attained a low current state (Figure 5a). Interestingly, this sample gained a rather stable low current state during the second MR study (Figure 5b). Immediately, after the observation of this low current state, after the second MR study, a current-voltage study was performed. Current-voltage data in the inset of figure 5a correspond to the stabilized low current state. It appeared that an in-plane magnetic field was assisting the stabilization of suppressed or low current state. This magnetic field induced low current state is in agreement with the other incidences of current suppression observed on the same magnetic tunnel junction configuration [61].

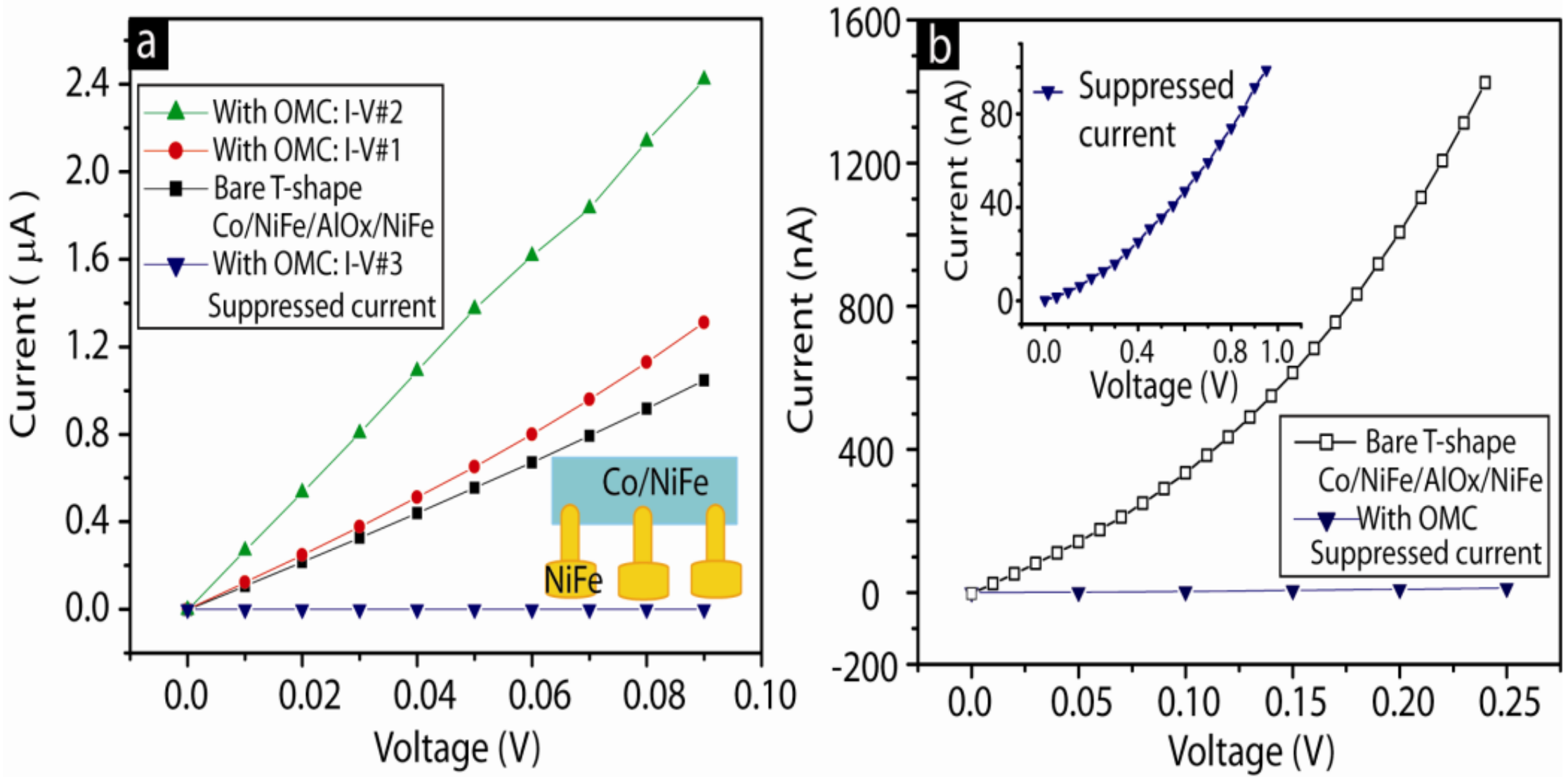

**Figure 6.** Current suppression with isolated TJMDs: (a) on isolated Co/NiFe/AlOx/NiFe tunnel junction OMCs first increased the overall current level, and eventually a suppressed current state stabilized. (b) Suppressed current state was stable. Inset graph show that current-voltage study in the suppressed current state up to 1.0 V.

The TJMD device reported in figure 5 is one of the many devices present in the cross bar or grid pattern on a chip [61]. It is important to ensure that noise and other unpredictable influences from neighboring tunnel junctions did not lead to spurious data which was perceived as OMC- molecule effect. To address this concern a single chip with a number of independent tunnel junctions and a broad bottom electrode was studied [61] (Inset of Figure 6a). This device configuration is important to avoid the effect of potential artifacts; the bottom metal electrode, which contains  Co, an etching prone metal in molecular solution, was protected by NiFe from the top in ~10 $mm^2$ area [58]. A broad bottom electrode ensured that any localized damage will not affect the tunnel junction current. Another advantage of this device geometry is that all the tunnel junctions are truly independent from each other, as long as their tunnel barrier is intact; tunnel barrier quality can be judged by the current–voltage study prior to the molecule attachment. One independent tunnel junction exhibited a current increase due to OMCs (Figure 6a). After multiple current-voltage studies, overall current settled in the suppressed current state (Figure 6b). This suppressed current state remained stable up to 1 V (Figure 6b). Several other cases and control experiments are discussed elsewhere [60, 61]. The observation of OMCs induced dramatic current suppression was supported by the equally dramatic OMCs induced changes in the magnetic properties of the magnetic tunnel junctions. In depth details of magnetic studies are furnished elsewhere [33] and succinctly within this paper.

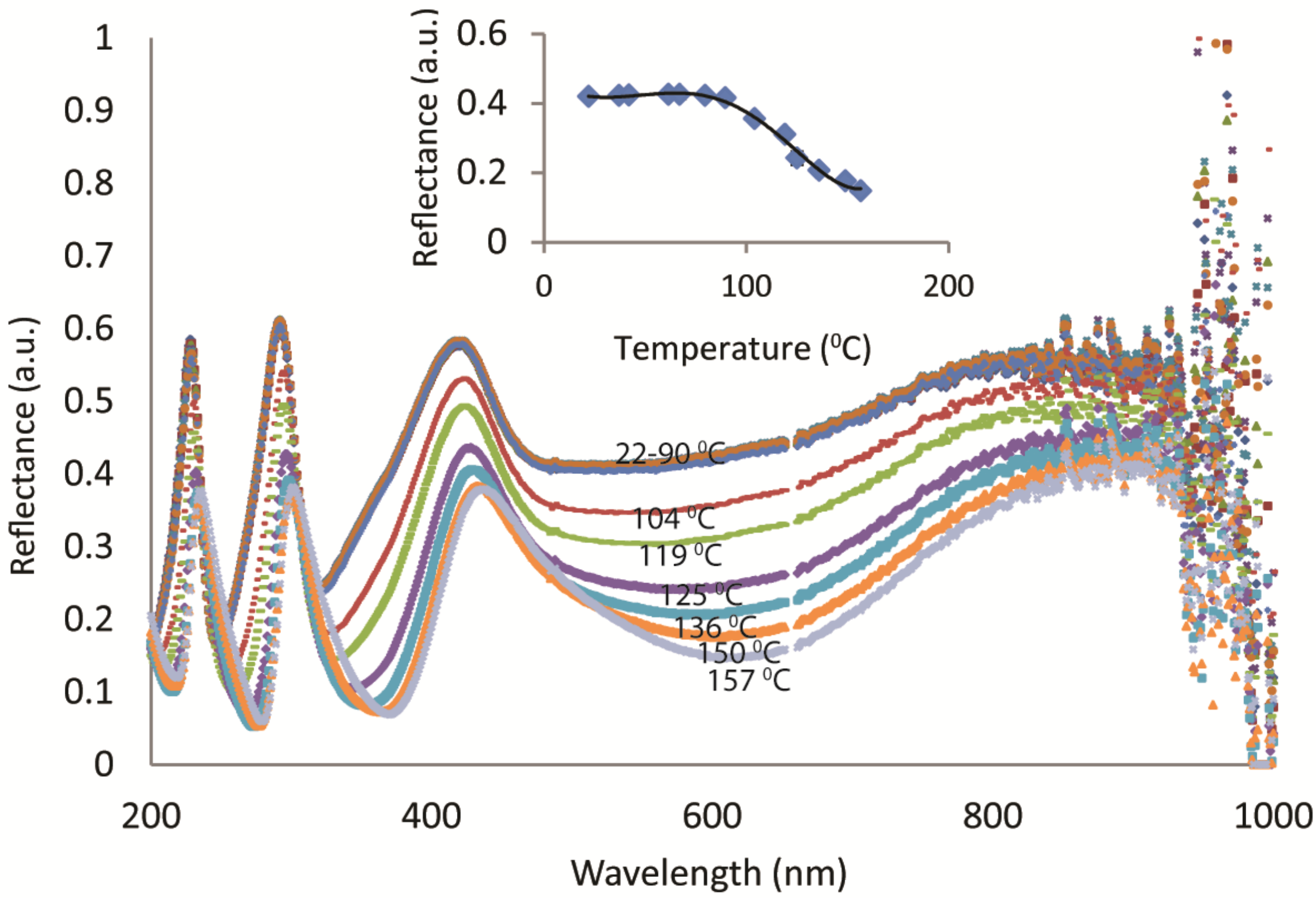

**Figure 7.** Spectral reflectance measurement on NiFe(10 nm) deposited on 2 nm Ta seed layer. Inset graph shows NiFe reflectance for 600 nm wavelength variation with temperature.

Criteria for selecting an arbitrary metal as a TJMD electrode must consider the stability against oxidation. It must be noted that even though one can make molecular device with almost any metal or semiconductor the electrode properties will keep changing with time if it has a tendency to oxidize. So far Au has been the most widely used electrode [3, 11, 28]. Other metals may even perform better than Au in molecular devices with regards to making chemical bonds with thiol linkage [62]. Recent experimental studies show that NiFe, along with being ferromagnetic, is a highly promising candidate as an electrode material. NiFe exhibited excellent oxidation resistance in low temperature range [59]. Furthermore, NiFe was used as an electrode in a ~5 nm thick $C_{60}$ molecule based spintronics devices [63].

To check the NiFe ability to withstand room temperature oxidation we performed electrodeposition of Au, and spectroscopic reflectance studies. A bilayer of 2 nm tantalum (Ta) and 10 nm NiFe was sputter deposited on an oxidized Si wafer with ~300 nm thick $SiO_2$. Hence, Ta only served as a seed layer [64, 65] to increase the adhesion of NiFe to substrate. Electrochemical Au deposition occurred on NiFe aged for more than one month at room temperature. It signifies that NiFe is stable in air. This experiment suggests that NiFe can electrochemically bind with thiol like terminal groups to make chemical bonds with the molecules of interest [9, 50, 66, 67]. Electrochemistry based metal-thiol reaction is found to be highly efficient and quick [49, 51].

Recently we studied the effect of temperature on the stability of ~10 nm thick NiFe deposited on ~2 nm Ta seed layer. We chose a small thickness as NiFe thickness in TJMDs is expected to be around 10 nm [18, 33]. Then Ta (2nm)/NiFe(10 nm) was heated from 22 $^0$C to 80 $^0$C temperature range; at every temperature setting a sample was heated for 30 minutes. NiFe still allowed electrochemical Au deposition when heated at 80 $^0$C; however, heating NiFe at 150 $^0$C caused visual changes in the film color. NiFe/Ta bilayer color changed from silver white to deep blue; it is experimentally clear that at 150 $^0$C ~ 10 nm NiFe and Ta completely oxidized. One direct corollary of this experiment is that during molecular device fabrication the NiFe metal electrode should not be exposed to air at temperature higher than 90 $^0$C.

It is noteworthy that during liftoff based TJMD fabrication approach, scheme that utilized NiFe most effectively for molecular device fabrication [58], briefly exposed bottom electrode's NiFe to air at 80-90 $^0$C temperature during the second photolithography stage. However, NiFe was covered by the photoresist during this baking step. To deposit AlOx insulator and on top of that NiFe electrode, photolithography was performed on oxidized Si with a preexisting patterned NiFe or Co/NiFe bottom electrode [18]. During this step the sample was spin coated with a thin ~500 nm photoresist layer and was baked in 80-90 $^0$C range for 45-60 seconds [58]. One critical reason for doing baking in 80-90 $^0$C temperature range was to have easy liftoff; the liftoff was the key step that must provide clean edges for molecules bridging. TJMDs produced in this manner showed clear effects of molecule attachments. The same protocol was adopted for the sample which showed molecule reversal effect (Figure 3c) [18]. More importantly TJMDs for the magnetic studies [33] were deposited through single photoresist cavity and were never exposed to air above room temperature. It is apparent that using NiFe metal in molecular devices without exposing to high temperature air played an important role in observing rather intriguing current suppression [60].

To further investigate the transition in NiFe properties with temperature spectroscopic reflectance studies were performed. These reflectance studies signify that significant oxidation starts around 90 $^0$C; the entire reflectance scan for less than 90 $^0$C overlapped (Figure 7). After 90 $^0$C NiFe film started showing a significant change in the reflectance property (Figure 7). Other studies which investigated a relatively thick NiFe film (180 nm) indicated that insignificant oxidation occurred below 300 $^0$C [59]. With these studies it is clear that the molecular device community can use NiFe as a magnetic electrode safely if they can avoid heating NiFe beyond 90 $^0$C in the presence of air. However, doing annealing to improve tunnel junction quality in inert ambience or vacuum should not affect NiFe capability to host molecules. Liftoff based TJMD fabrication approach [18, 33] has successfully demonstrated the use of NiFe as a promising electrode in a number of studies.

*3.6 TJMD based molecular spintronics*

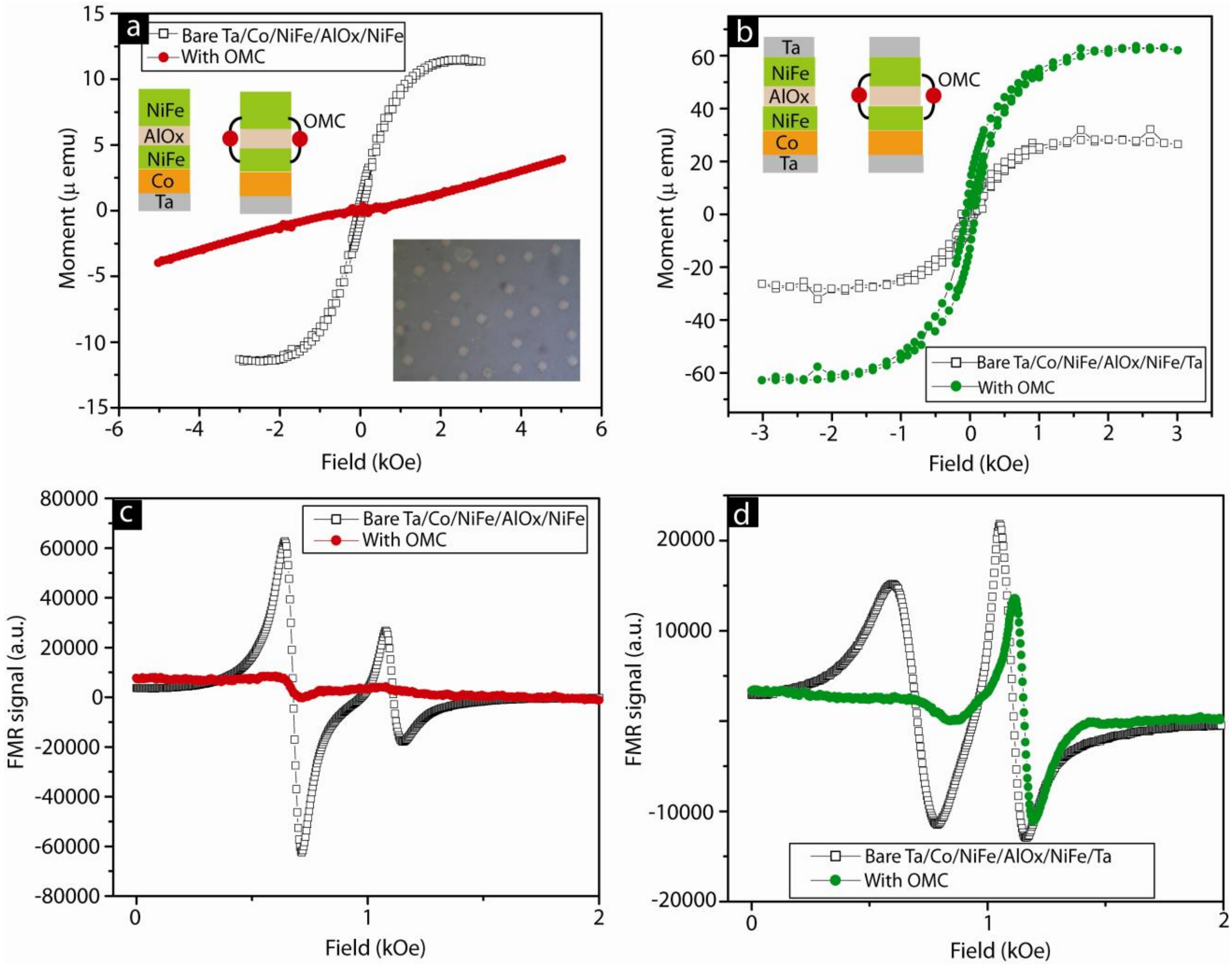

**Figure 8.** Effect of OMCs on the MTJ's magnetic properties: (a) OMCs made magnetization of Ta/Co/NiFe/AlOx dots linear. (b) OMCs increased the magnetization of Ta/Co/NiFe/AlOx/NiFe/Ta. OMCs strongly influenced the FMR of (c) Ta/Co/NiFe/AlOx/NiFe and (d) Ta/Co/NiFe/AlOx/NiFe/Ta.

Molecular spintronics devices are expected to show much higher tunneling magneto resistance as compared to their inorganic counterparts [68-71]. Molecular spintronics devices are also expected to be utilized in computational devices [9, 72]. Nobel laureate Dr. Fert referred to molecular spintronics devices as the promising memory device architecture for future [73]. Despite enormous fanfare the realization of commercially viable molecular spintronics device remains a daunting task. The TJMD approach, which is an extension of magnetic tunnel junction based devices [56], can be a practical solution to the molecular spintronics device fabrication issues. Magnetic tunnel junctions can be easily transformed into a molecular spintronics device; one only needs to use air stable ferromagnetic electrodes and suitable molecules [9, 57, 74] to realize such devices. One of the authors has contributed in the development of magnetic tunnel junction based molecular devices [33, 58, 61]. These devices were based on Co/NiFe/AlOx/NiFe magnetic tunnel junction and OMCs [50]. This combination yielded intriguing transport properties such as several orders of current suppression at room temperature due to OMCs [61, 75]. A large number of control experiments indicate that a current suppression phenomenon is associated with the OMCs enhanced magnetic coupling between the two ferromagnets [33]. Molecules bridging the gap between two ferromagnets appear to be much more than a simple transport channel [13, 69]; molecules appear to serve as the magnetic coupling agent which brings about transformational changes in the ferromagnetic films of a tunnel junction itself [14, 76, 77].

Interestingly, TJMDs are naturally suitable for magnetic studies to explain the effect of molecular device elements on the overall magnetic and transport properties of a bare magnetic tunnel junction testbed [56, 78-80]. One can easily perform SQUID magnetization, ferromagnetic resonance (FMR), magnetic force microscopy (MFM) etc. on a magnetic tunnel junction before and after transforming it into a molecular spin device [33]. For instance, the magnetic studies were conducted on the same combination of a magnetic tunnel junction (Co/NiFe/AlOx/NiFe) and OMCs [18, 50], which produced the current suppression phenomena at room temperature [33].

The SQUID magnetometer and Ferromagnetic resonance (FMR) studies exhibited OMCs dramatic effect on the magnetic properties of Ta/Co/NiFe/AlOx/NiFe and Ta/Co/NiFe/AlOx/NiFe/Ta tunnel junction (Figure 8). It must be noted that Ta on top played an important role in defining the nature of inter-ferromagnetic electrode coupling in Ta/Co/NiFe/AlOx/NiFe tunnel junction. According to FMR studies, inter-ferromagnetic electrode coupling was antiferromagnetic without a top Ta layer and ferromagnetic with a top Ta layer [33]. For the magnetic studies several thousands of the magnetic tunnel junction pillars with the exposed sides were employed; the idea behind using several thousands of magnetic tunnel junctions on the same chip was to enhance the magnetic signals during magnetic studies.

The SQUID magnetization study showed that Ta/Co/NiFe/AlOx/NiFe tunnel junction exhibited a typical magnetization loop prior to interacting with OMCs (Figure 8a). When OMCs were chemically bonded to the two ferromagnetic electrodes, the magnetic field versus magnetization curve turned into a straight line (Figure 8A). According to the literature, the evolution of a linear magnetization curve signifies the existence of a strong anti-ferromagnetic coupling between the two ferromagnetic electrodes [81]. However, Ta/Co/NiFe/AlOx/NiFe/Ta, which had ferromagnetic inter electrode coupling, exhibited an increase in magnetization after hosting OMCs (Figure 8b). Extensive details about the OMC effect on tunnel junctions with ferromagnetic electrode(s) are published elsewhere [33].

In another magnetic study, the FMR measurements were conducted before and after treating Ta/Co/NiFe/AlOx/NiFe magnetic tunnel junction with OMCs. The FMR studies on the bare Ta/Co/NiFe/AlOx/NiFe tunnel junction showed a weak antiferromagnetic coupling between the two ferromagnetic electrodes of this tunnel junction [82]. However, bridging OMCs between ferromagnetic electrodes dramatically changed the FMR response. Major peaks associated with Ta/Co/NiFe/AlOx/NiFe tunnel junction disappeared (Figure 8c). According to the theoretical calculation by Layadi, [83, 84] such ferromagnetic resonance response corresponds to the development of a very strong antiferromagnetic exchange coupling between the two ferromagnets. FMR of Ta/Co/NiFe/AlOx/NiFe/Ta tunnel junction was transformed by the OMCs as well [33]. Two FMR peaks associated with Ta/Co/NiFe/AlOx/NiFe/Ta tunnel junction became one (Figure 8d); according to the theoretical study for an analogous system strong ferromagnetic coupling between two ferromagnets make two electrodes behave like one ferromagnetic layer and produced one FMR peak [83].

Ta/Co/NiFe/AlOx/NiFe tunnel junction exhibited OMC induced current suppression (Figure 5 and 6) which is directly correlated with OMCs induced strong antiferromagnetic coupling (Figure 8a and c). Presumably OMCs, possessing a paramagnetic nature [50, 57], appear to cause spin filtering for the electrons traveling through them and finally leading to spin polarized electrodes. Petrov et al. [85, 86] theoretically suggested spin filtering and spin polarization due to a paramagnetic ion and molecule.

The spin filtering and spin polarization caused by a paramagnetic molecule appear to affect the spin density of states of the ferromagnetic electrodes. The spin polarization of ferromagnetic electrodes in a magnetic tunnel junction strongly depends on the nature of tunneling medium. For instance, a magnesium oxide tunnel barrier leads to a higher spin polarization than that of AlOx tunnel barrier [56, 80]. Spin polarization is directly associated with the density of states of spin in a ferromagnetic material. In another related study, reducing the nonmagnetic spacer between nickel ferromagnetic metal affected the Curie temperature like important properties [87]. In the present case, OMCs, after becoming a strong magnetic coupling bridge, are presumably increasing the spin polarization of the two ferromagnetic electrodes to influence their density of states. Affecting the density of states of ferromagnetic electrodes will affect the tunneling current through both AlOx insulating spacer [88, 89] and the molecular conduction channel (Figure 1d-e) [90].

In a related study, the interaction between a $C_{60}$ molecule and nickel (Ni) ferromagnetic electrodes exhibited intriguing and unprecedented Kondo level splitting [14]. This Kondo level splitting was in agreement with the theoretical predictions of exchange splitting on analogous system comprising of a quantum dot coupled ferromagnetic electrodes [76]. The exchange interaction between ferromagnetic electrodes via molecules was shown to produce a very strong local magnetic field leading to the Kondo level splitting; the magnitude of the molecule induced magnetic field was several tens of Tesla and believed to be more than that produced by manmade magnets. The molecule induced effect on Kondo level splitting may also impact the Ni electrode density of states. However, no magnetic study was performed to understand the magnetic state of Ni break-junction or Ni electrodes before and after hosting $C_{60}$ molecule within the break junction gap. Apparently a TJMD approach is suitable for enhancing magnetic signals simply by increasing device density and doing magnetic studies before and after molecule attachment.

The TJMD approach has a unique ability for developing molecular spin devices with several promising molecules like single molecular magnets [91, 92], porphyrin [67, 74, 93, 94], and DNA [66] etc. Magnetic tunnel junctions have been optimized in different laboratories and industries worldwide [56]; potential molecular spintronics technology getting developed through the TJMD approach will be commercialized in a relatively short time span. Initiating and strengthening the collaboration between chemists and the magnetic tunnel junction community can be highly profitable and may produce novel inventions.

*3.7 Economical fabrication and commercialization possibility*

A TJMD has strong potential to become commercially viable because it utilizes commonly available microfabrication tools [3, 95]. A TJMD approach also democratizes opportunities for the worldwide researchers. A TJMD's tunnel junction have been produced using photolithography and physical vapor deposition systems [18]; after that molecular self-assembly process commenced to establish molecular conduction channels as the final step to complete molecular device fabrication [3]. Photolithography and physical vapor deposition systems are an integral part of the commercial fabrication of the present day silicon devices and are also available in almost every lab involved in any form of device research. A TJMD can also become very economical as one can produce them with high-yield. TJMD can be integrated into the existing CMOS devices as they are produced using compatible device fabrication resources [8].

The utilization of magnetic tunnel junctions as the test bed in TJMDs will lead to even more exciting opportunities for developing novel devices [33, 75]. Magnetic tunnel junctions are used as an important component in the read heads of present day computer’s memories [56]. The development of magnetic tunnel junction based molecular devices can be boosted by the utilization of commercially successful or promising magnetic tunnel junction configurations [80, 96].

*3.8 Cross bar and stackable device structure*

The cross bar junction design is highly suitable to densely pack molecular devices [97-100]. A cross bar design is especially appropriate for spintronics devices [79]. A simple photolithographically defined cross bar TJMD is suitable for realizing individually addressable cross bar device architecture [18, 58]. The yield and success of this scheme will depend on the process control parameters to achieve robust and long lasting tunnel junction test beds in a cross bar pattern [46]. There are possibilities that many tunnel junctions may be short circuited before, during, and after the interaction with molecular device elements. These issues can be fixed by careful process optimization. It is noteworthy that with a TJMD approach all the tunnel junctions can be electrically tested before and after the inclusion of molecular device elements to determine which tunnel junction got transformed into a molecular device [3, 18, 33]. Other device fabrication approaches discussed in section 2 are not well suited for such cross bar like device architecture; it may be very challenging to adopt them for the commercial molecular device fabrication.

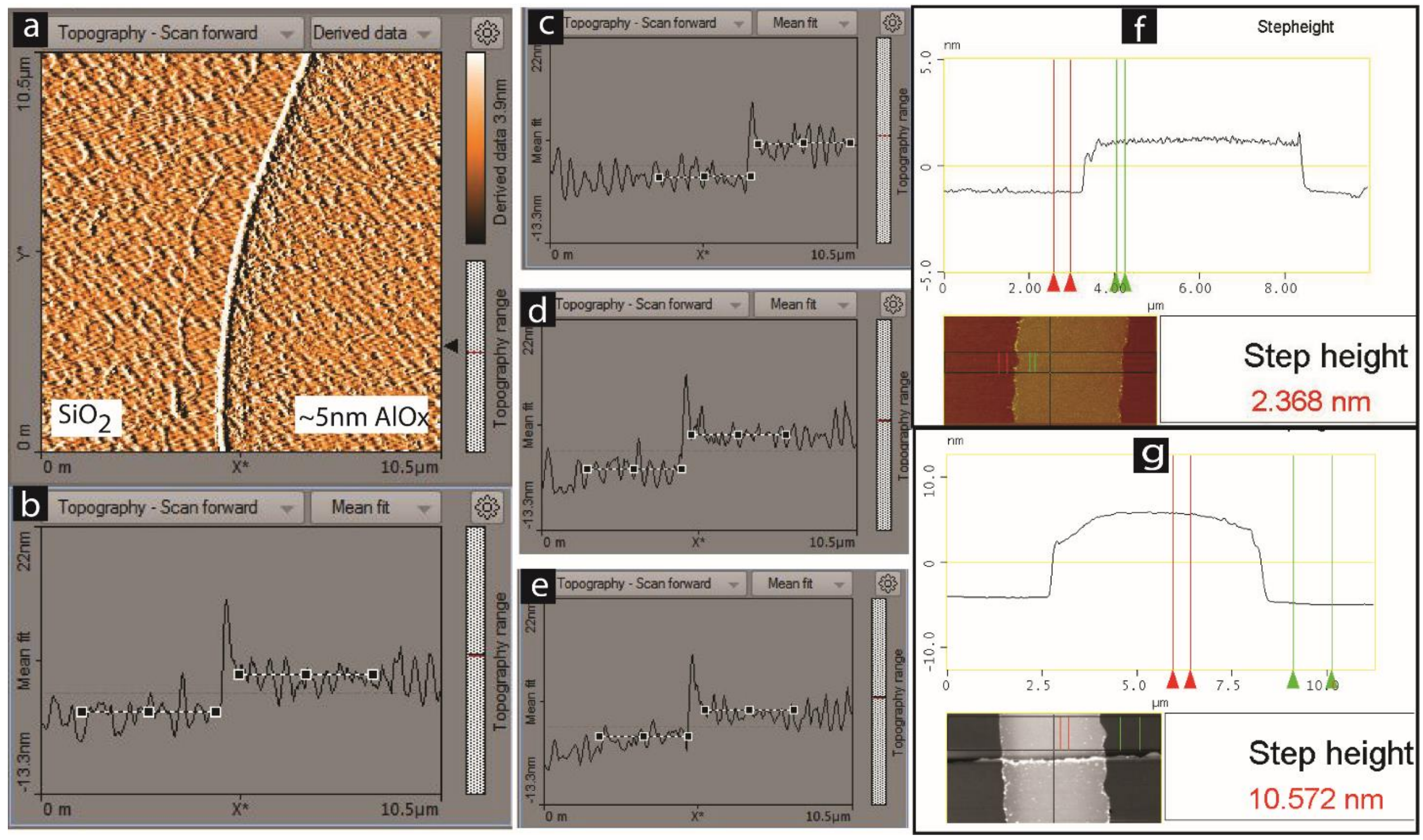

**Figure 9.** Evolution of tall notches at the edges of AlOx insulator: (a) Top view of a sputter deposited ~ 5 nm AlOx in unoptimized photoresist cavity profile produced (b-e) notches along the edges. (f) Low pressure AlOx deposition without rotating sample can produce very low edge notches. (g) Tapered edge profile can be achieved by developing an undercut in photoresist cavity.

Additionally, TJMDs are stackable for realizing numerous unexplored forms of dense molecular device circuitries. Multiple tunnel junctions with at least one exposed side edge can be vertically stacked; these multiple tunnel junctions in a stack can have different electrode materials and spacer thickness to allow the realization of many types of TJMDs. Different TJMDs on the same chip can simultaneously perform different functions of (i) logic and memory devices, (ii) electronic and spintronics device; (iii) optoelectronics and electronics device etc. This attribute is unique to a TJMD device approach. Other forms of molecular devices discussed in section 2 may not be stacked in practical manner.

## 4. Limitations and challenges of TJMD approach

The TJMD approach also has a number of limitations and issues. Fortunately, many of them can be addressed with the appropriate process optimization and do not lead to a dead end. To enhance the longevity of a TJMD it is critical to optimize tunnel barrier stability [46, 101, 102]; it was noted that many tunnel junction testbeds possessed process induced mechanical stresses which led to an evolution of defects within the tunnel barrier [46]. It will be important to keep the mechanical stresses in check while producing TJMDs for the commercial and long-term applications.

The thickness of an insulator barrier at the edges of a tunnel junction testbed is also critical. The height of insulator along the tunnel junction edge strongly depends on how a tunnel junction is fabricated. Generally, this is not the issue with an ion milled or chemically etched tunnel junctions where one can get the sharp vertical edges. However, etching and milling leads to device instability due to stress generation. A simple photoresist liftoff process can avoid the need for doing ion milling and chemical etching for creating inter-electrode separation equal to tunnel barrier thickness [18, 33]. The liftoff based approach, which uses photolithography to produce exposed side edges of the tunnel junction, can lead to notches at the edges (Figure 9a). Due to this the insulator thickness may be higher at the edges than the thickness of the insulator in the interior [3] (Figure 9a-e). Figure 9b-e exhibits multiple AFM images of steps along AlOx edge showed in Figure 9a. This issue can be solved by using a proper photoresist profile and appropriate deposition conditions [18]. Depositing AlOx insulator in ~500 nm thick photoresist thickness with a static sample position with respect to the sputtering gun produced negligible notches along the insulator edges (Figure 9f). Importantly, it is advisable to keep sufficient gap between the sputtering gun and a sample, and low sputtering gas (argon) pressure to achieve a quasi-line of sight deposition in the sputtering machine as well. One can also use photoresist cavities with undercut profile to achieve even tapered edges (Figure 9g). Producing an AlOx with tapered edges will yield higher AlOx thickness at the center and can provide low leakage current via insulator barrier to start with.

A brief chemical etching step for the insulator barrier only can also easily remove the notches along the edges of insulating barrier [3]. For instance, in order to remove AlOx tunnel barrier near tunnel junction edges one can simply use a dilute NaOH solution; it is a common practice in template based nanowire fabrication to dissolve ~50 µm thick AlOx template without damaging metallic nanowires [103, 104].

In addition, the liftoff based TJMD fabrication approach is likely to suffer with sharp notches along the bottom electrode; the bottom electrode for this scheme is strongly advised to be produced by depositing in a photoresist cavity with an undercut profile. The undercut profile in photoresist successfully produced tapered edges on the bottom electrode (Figure 9g).

To resolve the notch issues at the edges on the bottom electrode and the insulator the use of shadow mask may be highly useful. A shadow mask can be used when depositing the bottom electrode, insulator and the top electrode in a TJMD and this will avoid the need of doing photolithography. The usage of a shadow mask may also bring additional advantage. Due to the shadow mask the oxidation sensitive metal electrodes will not encounter high temperature baking during photolithography. This recommendation may be especially helpful for the less experienced research groups willing to produce TJMDs. Incorporating a shadow mask technique can shorten the process optimization dramatically. However, there may be a need to perform very brief insulator etches, either chemically or by milling to ensure that tunnel barrier dimensions are not greater than the top electrode dimensions in the junction area; by this way approach the inter-electrode gap will be equal to the insulator thickness.

In the TJMD approach the molecular attachment process occurs after the completion of the tunnel junction fabrication. One must be careful about checking if the molecular solution or protocol used for molecular self-assembly step is going to damage components of the tunnel junction. Some commonly used organic solvents like dichloromethane can start having acid content over a period of time and that can attack some of the metals of the tunnel junction testbed. For instance, OMC solution in dichloromethane attacked Co but did not affect NiFe [58]. Due to the ability to withstand chemical etching NiFe [58] was utilized as a protective layer on the Co layer in previous TJMDs [33]. Advantageously, Co/NiFe bilayer was smoother and has different magnetic hardness as compared to NiFe only.

There is another set of more challenging limitations for the TJMD approach. Integration of a gate electrode is quite challenging with the TJMD device architecture. This may be a major problem for the commercialization of the TJMD approach as the logic devices. However, this issue may not be that important for the TJMD design based molecular spintronics devices where the electron's spin property would play the key role.

Designing a TJMD with a single molecular device element is also challenging. Limiting the area of a TJMD's exposed side edges to ~ 1 nm only to accommodate single molecules is very difficult for the commercial fabrication capabilities. The potential but, costly solution to this problem may be to nanolithographically pattern the area to host the one or very few molecules [105, 106]. However, for future commercial applications one may not be interested in the single molecular TJMD devices. Multiple molecular device elements can make the overall device performance robust and defect tolerant; if one molecular connection is faulty then other properly connected molecules can take care of device functionality.

A TJMD is not at all suitable for molecules that are not designed to make chemical bonding with the metal electrodes. Molecules like $C_{60}$ are extremely challenging to position between the two metal electrodes on the side edges of a tunnel junction testbed. They are better studied with the metal break junction type devices [14, 15].

## 5. Future applications of TJMD

A TJMD approach can provide novel logic devices, memory devices, bio sensors, microwave sensors, spinphotovoltaic cells etc. A TJMD is a natural candidate for the applications where magnetic tunnel junctions (MTJs) or tunnel junctions play a central role.

MTJ based TJMD can be used for developing logic devices. It is noteworthy that MTJ-based programmable logic devices have been considered for reconfigurable and nonvolatile computation devices [107]. A TJMD can make even better use of spin transfer torque (STT)-based switching. STT has unique advantages in device scaling compared to the field effect based -switching mechanism [8]. New device architecture will be crucial for developing TJMD-based logic devices. These TJMDs based logic unit can function using the STT switching mechanism [107]. Such TJMDs can also be used as a means to reduce the standby power losses due to high leakage current in scale down CMOS circuits [108]. Similar to MTJs, TJMDs can be fabricated to establish direct communication amongst them to yield efficient logic operation [109].

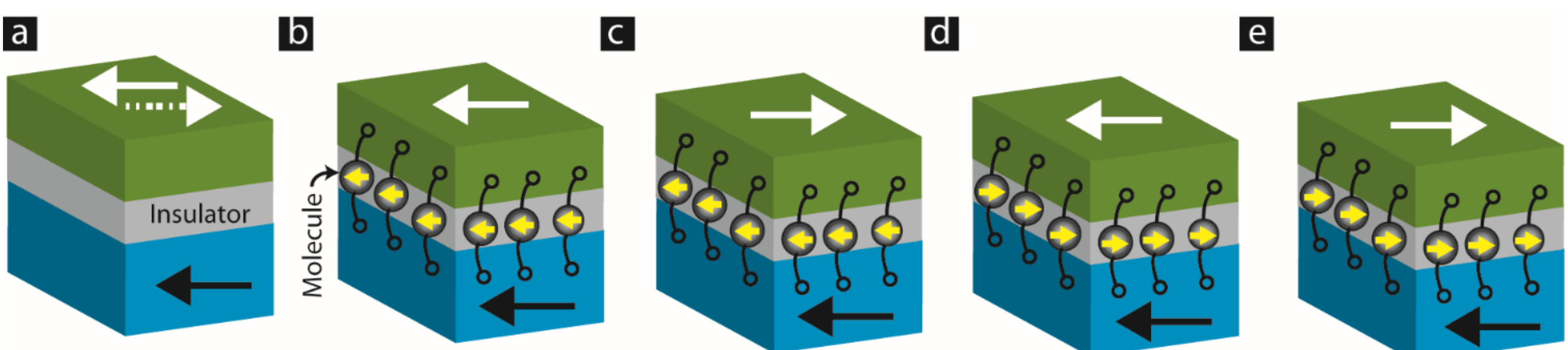

**Figure 10.** (a) Magnetic tunnel junction with pinned bottom magnetic layer and free magnetic layer at the top. Magnetic tunnel junction with paramagnetic molecules may exist in the states where (b) spins of molecule, top electrode, bottom electrode in the same direction, (c) spin of the top electrode is opposite to the spins of molecules and bottom electrode, (d) spins of molecules are opposite to the spins of top and bottom electrodes, (e) spins of the top electrode and molecules are opposite to the spin of bottom electrode.

A TJMD approach is highly suitable for developing magneto resistance devices for readheads and Magnetic Random Access Memory (MRAM) application [56]. The TJMD material stack includes two magnetic layers separated by an ultrathin dielectric barrier with molecular conduction channels around its edge (Figure 1d-e). A TJMD based MRAM will rely on a device mechanism that pin the polarization of one of the magnetic layers in a fixed direction and keep the other ferromagnet free to rotate. The polarization direction of the free magnetic layer can be used for information storage. The resistance of the memory bit is expected to be either low or high, depending on the relative polarization (parallel or antiparallel) of the free layer with respect to the pinned layer and molecular spin. An applied field can switch the free layer and molecule spin with respect to pinned layer magnetization direction. A MTJ only has two resistance states due to parallel and antiparllel magnetization of its two magnetic electrodes (Figure 10a). A TJMD based MRAM may have four or more magnetic states or resistance sates (Figure 10b-e). Similar to MTJ based MRAM array, in TJMD orthogonal lines can pass under and over the bit, carrying current that produces the switching field. According to theoretical studies molecules connected between two ferromagnetic electrodes can yield very high magneto resistance ratio [69, 110]. A TJMD approach can utilize various MTJ configurations as the testbed and improvise its magneto resistance properties by bridging promising molecular channels. A TJMD can also benefit from the industrial processes used for making MTJ read heads [56]; in one sense the TJMD technology is already partly commercialized. Hence, among all the other molecular device fabrication approaches, the TJMD approach is closest to commercialization or mass production. A TJMD is capable of using stellar attributes of molecular device elements, such as low hyperfine splitting, low spin –orbit coupling, and high spin coherence time and length [70].

A MTJ based TJMD device can also be used as a sensor for microwave energy. It is noteworthy that MTJs have exhibited change in tunneling magneto resistance in the presence of external microwaves [111]. The variation in the relative angle between the two ferromagnetic electrodes and molecular device element's spin may result in a rectification of the average resistance change. Scalable MTJ based TJMDs have the potential to exhibit very high sensitivity to the microwave-magnetic field, and hence can serve as a sensor for microwave power and microwave frequencies.

A MTJ based TJMD can produce novel magnetic metamaterials. Molecular device channels can be utilized to change the magnetic coupling between the two ferromagnetic electrodes. Molecules can enhance the magnetic coupling to the extent that overall device can start exhibiting entirely different magnetic properties with respect to a bare magnetic tunnel junction [33]. Several theoretical studies discussed the expected magnetic properties from strongly coupled ferromagnetic electrodes [83, 84]; TJMD approach has strong potential to test these predictions and initiate a new approach of creating magnetic meta-materials.

A MTJ based TJMD can enable the observation of spin galvanic effect. Molecules between the two ferromagnets are akin to quantum wires. Fedrove et al. [112] theoretically showed the current generation in a quantum wire by the external electromagnetic radiation. The photocurrent is believed to arise from the interplay of spin-orbit interaction and an external in-plane magnetic field. A MTJ based TJMD will be uniquely suitable to investigate such novel concepts.

A TJMD can function like an optoelectronic or opto-spintronic device. By the virtue of TJMD's design molecular conduction channels are present at the exposed edges and can be exposed to light radiation. Different molecules can respond to different light wavelengths to facilitate the emergence of novel optoelectronics and opto-spintronics devices. One can also study the optical properties of individual molecules connected to metal electrodes.

The TJMD scheme has tremendous opportunity to serve as a biosensor. Several components of a TJMD including the electrodes and molecules in the exposed regions, can interact with one or more chemicals or biochemicals [113]. Utilization of Au or platinum as metal electrodes may enable a TJMD to do electrochemistry based neurochemical sensing [114], while molecular conduction channels may perform field effect based sensing of other neurochemicals. In essence a TJMD can be used to do electrochemical and field effect based sensing with extremely high sensitivity and spatio-temporal resolution. In addition, molecular device channels can be specifically designed to selectively bind with the desired molecules and enable field effect based sensing [115]. The utilization of molecule as a sensing element can be used to target the miniscule amounts of hazardous molecules and polluting gases. If MTJs are used in a TJMD then the magnetic effect will enhance its utility as a unique and highly versatile sensor [116]. One can target the detection of DNA with high sensitivity [117]. The tunneling magneto resistance phenomena on a TJMD may be very sensitive towards the chemical ambience; molecules can be specifically designed not only to serve as conduction channels but also to interact with chemicals and bio-chemicals of interest. The TJMD approach opens up new opportunities to do chemical sensing in an unprecedented manner.

## 6. Conclusions

This paper discussed the TJMD approach as a potential route to develop molecule based devices. A TJMD approach is amenable to a large number of control experiments, including reversal of molecular conduction channels. A TJMD approach utilizes a prefabricated tunnel junction as the test bed, and converts it into a molecular device simply by bridging the molecules of interests between two electrodes in the exposed regions. Fabrication of tunnel junctions is commercially viable and compatible with CMOS technology. Hence, a TJMD approach is much closer to commercialization as compared to any other molecular device fabrication approach. One of the biggest advantages of the TJMD approach is its ability to incorporate virtually any metallic or semiconducting electrode. A TJMD can utilize ferromagnetic electrode. This paper recommended safe temperature limit and a device fabrication protocol to minimize the possibility of oxidation, alternatively vacuum and inert environment should be used during any processing step at temperature more than 80 $^{o}$C. Unfounded apprehensions and overestimation of oxidation impact on ferromagnetic electrodes has deterred the development of molecular spintronics devices [118].

A TJMD with ferromagnetic electrodes and paramagnetic molecules have demonstrated several orders of current suppression. This observation of current suppression is supported by the molecule induced strong magnetic coupling between the two ferromagnetic electrodes of the same magnetic tunnel junction and molecule combination that showed current suppression. A TJMD approach is unique in allowing magnetic studies to understand the behavior of molecular spintronics devices. Bright prospects of realizing novel molecular spintronics devices by utilizing ferromagnetic electrodes and paramagnetic molecule are also apparent from the nickel breakjunction study. Previously, Pasupathi et al. [14] observed unprecedented Kondo level splitting with ferromagnetic break junctions. A TJMD approach relies on simple device fabrication scheme and facilitates a large number of control experiments and magnetic studies to understand molecule behavior. Further investigations by independent research groups are in order to confirm and understand the effect of molecular conduction bridges leading to current suppression and inter-ferromagnetic electrode coupling in TJMDs. This paper invites potential researchers to study other forms of TJMDs for advancing current state of the art in molecular spintronics devices. This paper has discussed a number of experimental insights which may remove or reduce a number of crucial obstacles prohibiting the experimental studies.

The chemistry researchers apt in synthesizing molecules like porphyrins, single molecular magnets, organometallic complexes etc. will greatly enhance the TJMD research by teaming up with magnetic tunnel junction researchers. The collaboration between chemistry and magnetic tunnel junction researchers may also produce unique TJMDs for several high demand applications like novel logic and memory devices, biosensors, and electromagnetic energy sensors. A TJMD approach also creates opportunities to investigate exotic theoretical concepts like spin galvanic effect [112] and Barry phase oscillation [119].

**Acknowledgement**

We thankfully acknowledge funding support from the National Science Foundation (Award # HRD-1238802), Air force office of sponsored research (Award #FA9550-13-1-0152), and the US Department of Energy-National Nuclear Security Agency. Pawan Tyagi thanks Dr. Bruce Hinds and Department of Chemical and Materials engineering at University of Kentucky for facilitating experimental work on tunnel junction based molecular devices during PhD. Molecules for molecular device fabrication were produced Dr. Stephen Holmes's group. This paper does not necessarily represent the views of past and present affiliations of authors and funding agencies.

**Author Contributions**

This paper was written by Pawan Tyagi. Edwards Friebe studied NiFe oxidation using electrochemical deposition and spectroscopic reflectance study. Collin Baker studied NiFe oxidation with temperature using STM and performed thickness measurement on sputter deposited AlOx. Pawan Tyagi, Collin Baker, and Edwards Friebe refined the manuscript.